# Deterministic Compressive Sampling for High-Quality Image Reconstruction of Ultrasound Tomography

Tran Quang-Huy[1], Tran Duc-Tan[1], Huynh Huu Tue[2], Nguyen Linh-Trung[1]
[1]Faculty of Electronics & Telecommunications, VNU University of Engineering & Technology, Hanoi, Vietnam
[2]VNU International University, HoChiMinh, VIETNAM
E-mail: tranquanghuy @hpu2.edu.vn, tantd@vnu.edu.vn, hhtue@hcmiu.edu.vn, linhtrung@vnu.edu.vn

*Abstract*— A well-known diagnostic imaging modality, termed ultrasound tomography, was quickly developed for the detection of very small tumors whose sizes are smaller than the wavelength of the incident pressure wave without ionizing radiation, compared to the current gold-standard X-ray mammography. Based on inverse scattering technique, ultrasound tomography uses some material properties such as sound contrast or attenuation to detect small targets. The Distorted Born Iterative Method (DBIM) based on first-order Born approximation is an efficient diffraction tomography approach. Compressed Sensing (CS) technique was applied to the detection geometry configuration of ultrasound tomography as a powerful tool for improved image reconstruction quality. However, this configuration is very difficult to implement in practice. Inspired of easier hardware implementation of deterministic CS, in this paper, we propose the chaos measurements in the detection geometry configuration and the image reconstruction process is implemented using $L_1$ regularization. The simulation results of the proposed method have demonstrated the high performance of the proposed approach, the normalized error is approximately 90% reduced, compared to the conventional approach. Furthermore, with the same quality, we can save half of number of measurements and only use two iterations when using the proposed method.

## I. INTRODUCTION

Biomedical imaging technology has been making dramatic changes in the clinical diagnostic field. The explosive growth of the digital media and information technology offers the very clever and sophisticated methods for diagnosis and treatment [1]. In 1885, Wilhelm Roentgen discovered the X-ray beam, since then, the biomedical imaging technology was born. More than a hundred years, the development of advanced technology, from X-ray to MRI (Magnetic Resonance Imaging), CT (Computed Tomography), PET (Positron Emission Tomography), SPECT (Single Photon Emission Computed Tomography), UT (Ultrasound Tomography), EPR (Electron Paramagnetic Resonance), SWUI (Shear-Wave Ultrasound Imaging) and so on, has created large changes in clinical medicine. The effectiveness of non-invasive imaging modality rapidly developed with advances in computer science. Implementation ability for diagnostic and therapeutic procedures increases the widespread use of ultrasound.

Currently, ultrasound imaging techniques have become the most popular tools in the health sector, because of the advantages such as low cost, non-invasive nature, painless test, mobility and fast diagnosis.

Ultrasound imaging which uses sound waves in the range between 20 kHz and 1 GHz is commonly used since the development of sonar in 1910. Based on the principle of sonar, one of the techniques that can widely be used is B-mode imaging [2]. This technique is used for non-destruction evaluation and biomedical imaging. B-mode image represents a qualitative change of acoustic impedance function. Thanks to this change, it allows to distinguish different environments in the region of interest. However, this imaging technique, using feedback of sound waves when encountering target, only provides the qualitative information of the imaged targets. Meanwhile, ultrasound tomography, based on inverse scattering technique, provides the quantitative information of those targets.

Indeed, when sound waves encounter a heterogeneous environment, some of the energy will then be scattered in all directions. The scattered data will be obtained by the receivers which are set up around the target of interest. Therefore, a set of measurements of the scattered field is obtained. Inverse scattering problem includes estimating the distribution of acoustical parameters (such as speed of sound, attenuation and density) to reconstruct the target of interest in the inhomogeneous environment. This technique allows a more detailed description of the imaged target. Instead of using acoustical impedance parameter in B-mode imaging, it uses one of parameters of acoustical properties. Therefore, acoustic tomograms display quantitative information of the target under examination.

Although ultrasound tomography has many advantages, but this technique has not widely been applied in practice. One of the reasons is the lack of applications that can take advantage of inverse scattering techniques. Currently, the main application of this technique is only for breast imaging in women to detect cancer-causing cells [3-5]. Another limitation of inverse scattering techniques is lack of efficient and powerful calculation methods. Inverse scattering techniques have high computational complexity and it is also the main reason that there is so far a certain number of commercialized tomography devices. Hence, state-of-the-art inverse scattering techniques focus primarily on reducing the computational complexity and constantly improving the quality of imaging. Most of research works on ultrasound tomography are based on Born approximation. Born Iterative Method (BIM) and Distorted Born Iterative Method (DBIM) are well-known for diffraction tomography [6]. The DBIM is a quantitative approach in image reconstruction of the very small target. In this method, the background medium is considered inhomogeneous and is updated with each iteration. Therefore, the equation for Green's function and the equation for incident field are updated with each iteration.

Compressed sensing (CS), which is introduced by Candes and Tao [7] and Donoho [8] in 2006, could acquire and reconstruct sparse signals at a rate lower than that of Nyquist. Random measurement approach in the detection geometry configuration is proposed in [9]. A set of measurements of the scattered field is performed using sets of receiver's random positions. This method can reduce the computational complexity and improve the quality of the reconstruction of the sound contrast, compared to the linear measurement method. However, this method does not denoise well and is difficult to set up in practice. In [10], the authors proposed to use a chaotic measurement matrix, which is deterministic, instead of random one. Elements of the logistic sequence are generated by deterministic chaotic system which is so nonlinear, hence becomes random-like. The simulated results indicated that the chaotic approach outperformed the random approach in terms of the probability of exact reconstruction. Moreover, using chaotic CS system also inherits a simpler hardware implementation, compared to the random one. In this paper, we propose a method to enhance the reconstruction quality of ultrasound tomography by using the chaotic sampling technique in the detection geometry configuration. As a result, this approach will offer a very high performance, compared to the conventional DBIM method.

The paper is organized as follows. Section 2 presents the distorted Born iterative method and its algorithm. Section 3 describes the principle of compressive sampling, and then presents the fundamental of chaotic compressive sampling. Section 4 presents the proposed method (CCS-DBIM). Simulation results to illustrate the effectiveness of the proposed method in terms of normalized image reconstruction error are presented in Section 5. Finally, Section 6 concludes the paper and gives discussions on the proposed methods.

## II. DISTORTED BORN ITERATIVE METHOD

Fig. 1 illustrates the setting of transmitters and receivers around the target. The pressure signal from a transmitter is propagated, scattered and measured by receivers. The measured data would be brought to DBIM in order to estimate the sound contrast. The change of the sound speed would be utilized to detect any tissue if exists.

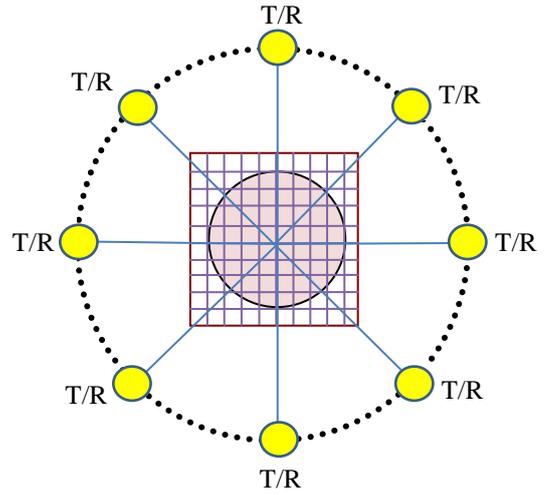

Fig. 1. Geometrical and acoustical configuration

There is a target whose density is a constant, placed in a homogeneous environment (e.g. water) of an infinite space. The wave numbers of the background and target mediums are $k_0$ and $k(r)$ respectively. The wave equation of the scheme can be expressed by:

$$p(\vec{r}) = p^{inc}(\vec{r}) + p^{sc}(\vec{r}), \qquad (1)$$

where $p^{sc}(\vec{r})$, $p^{inc}(\vec{r})$, and $p(\vec{r})$ are the scattered, incident, and total signals respectively.

Eq. (1) can be rewritten in details using the Green function $G_0(\cdot)$:

$$p(\vec{r}) = p^{inc}(\vec{r}) + \iint T(\vec{r})p(\vec{r'})G_0(k_0, |\vec{r}-\vec{r'}|)d\vec{r'}, \qquad (2)$$

When the background medium is homogeneous, $G_0$ is the 0-th Hankel function of the first kind

$$G_0(k_0, |\vec{r}-\vec{r'}|) = \frac{-i}{4}H_0^{(1)}(k_0|\vec{r}-\vec{r'}|) = \frac{-i}{4}\sqrt{\frac{2}{\pi k_0|\vec{r}-\vec{r'}|}}e^{i(k_0|\vec{r}-\vec{r'}|-\pi/4)}. \qquad (3)$$

T(r) in Eq. (2) is the target function that needs to be estimated. It can be calculated as follows:

$$T(r) = \begin{cases} k(r)^2 - k_0^2 = \omega^2\left(\frac{1}{c^2} - \frac{1}{c_0^2}\right) & if \ r \leq R \\ 0 & if \ r > R \end{cases} \qquad (4)$$

Eq. (4) indicates that the ideal target function depends on the frequency of the incident signal ($\omega = 2\pi f$) and the sound speed difference of the background medium ($c_0$) as well as the target medium (*c*). The method of moments (MoM) is used to discretize Eq. (2). Firstly, the total pressure field in the observed mesh area ($N \times N$ points) can be expressed by:

$$\bar{p} = (\bar{I} - \bar{C}.D(\bar{T}))p^{inc}, \qquad (5)$$

where $\bar{C}$ is the Green matrix showing the interactions among pixels, $\bar{I}$ is unit matrix, and $D(\cdot)$ returns a square diagonal matrix of the input vector. The scattered signal in form of $N_tN_r \times 1$ vector is described by

$$\bar{p}^{sc} = \bar{B}.D(\bar{T}).\bar{p}, \qquad (6)$$

where $\bar{B}$ is the Green matrix showing the interaction of all pixels to the receiver. We have to determine two parameters $\bar{p}$ and $\bar{O}$ in Eqs. (3) and (4). By rewriting these equations, we have [11]:

$$\Delta p^{sc} = \bar{B}.D(\bar{p}).\Delta\bar{T} = \bar{M}.\Delta\bar{T}, \qquad (7)$$

where $\bar{M} = \bar{B}.D(\bar{p})$. For a transmitter and a receiver, we formulate a matrix $\overline{M}$ and a scalar value $\Delta p^{sc}$. The target function $\bar{T}$ has $N^2$ variables corresponding to the number of pixels in the region of interest. It can be estimated by:

$$\bar{T}^n = \bar{T}^{(n-1)} + \Delta\bar{T}^{(n-1)}, \qquad (8)$$

where $n$ and $n-1$ are two consecutive discrete-time points. $\Delta\bar{T}$ is estimated by using Tikhonov's regularization [12]:

$$\Delta\bar{T} = \arg\min_{\Delta\bar{T}} \left\| \Delta\bar{p}^{sc}_t - \overline{M_t}\Delta\bar{T} \right\|_2^2 + \gamma \|\Delta\bar{T}\|_2^2 \qquad (9)$$

where $\Delta\bar{p}^{sc}$ is the difference between estimated and measured scattered signals whose size is ($N_tN_r \times 1$); measurement results are assembled in a matrix form $\overline{M}_t$ of ($N_tN_r \times N^2$) elements; $\gamma$ is the regularization factor that needs to be chosen carefully because it affects mostly to the stability of the system. High values of $\gamma$ make the reconstructed image rough. However, small values of $\gamma$ will lead to highly computational complexity.

The DBIM procedure is presented in Algorithm 1.

| Algorithm 1. The Distorted Born Iterative Method - DBIM |
|---|
| Set up the measurement configuration of linear tranmitters and receivers locations |
| Choose initial values: $\bar{T}_{(n)} = \bar{T}_{(0)}$ and $\bar{p}_0 = \bar{p}^{inc}$ using (21) |
| **For** $n = 1$ to $N_{DBIM}$, **do** |
| 1. Calculate $\bar{B}$ and $\bar{C}$ |
| 2. Calculate $p, \bar{p}^{sc}$ corresponding to $\bar{T}_{(n)}$ using (5, 6) |
| 3. Calculate $\Delta\bar{p}^{sc}$ using (7) |
| 4. Calculate $\Delta\bar{T}_{(n)}$ using Tikhonov regularization (9) |
| 5. Calculate $\bar{T}_{(n+1)} = \bar{T}_{(n)} + \Delta\bar{T}_{(n)}$ |
| **End For** |

## III. CHAOTIC COMPRESSIVE SAMPLING

### 3.1 Fundamentals of Compressive Sampling

Compressive Sampling (CS), also known as Compressed Sensing [13], allows exactly recovery signal $v \in \mathbb{R}^n$ from a small number of random measurements $u \in \mathbb{R}^m$ m < n. Random measurements $u$ may be collected in "sampling basis" $\Phi$, it depends on the collecting equipment. For example, in MRI, $\Phi$ is the Fourier basis. In ultrasound, $\Phi$ is simply common delta function, we have:

$$u = \Phi v, \qquad (10)$$

where $\Phi$ is a m×n matrix. The columns of $\Phi$ have entries (equal to 1) at random positions and zero in other positions, so the model randomly selected measurements.

The core problem of compressive sampling is that assuming $v$ has sparse representative in an orthonormal basis $\Psi$, ie:

$$v = \psi w \qquad (11)$$

In which, $w$ only has s < m < n non-zero coefficients. Signal $w$ is called sparse. Compressive sampling theory shows that this sparse property allows accurate recovery $w$ with overwhelming probability to matrix $\Phi\Psi$ [14]. In particular, sensing basis must have incoherent property to the model basis $\Psi$ [15]. This property is guaranteed by the randomness of the non-zero components in $\Phi$. Therefore, the problem can be written as follows:

$$u = \Phi\psi w = Aw, \qquad (12)$$

where A is a m×n full-rank matrix (i.e. the m rows of A is independent).

By these settings, the problem of CS is solving (12) for $w$, with $w$-sparse constraint. Once $w$ is solved, $v$ can be calculated from (11).

Matrix A, with a specified isometric constant which is called Restricted Isometry Property (RIP). Candes et al. [14] indicates that CS problem can be solved through $\ell_0$-minimization problem $P_0$:

$$\hat{w} = \arg\min_{w \in \mathbb{R}^n} |w|_{l_0} \text{ subject to } u = Aw, \qquad (13)$$

in which, $l_0$ norm is $\|w\|_{l_0} := |\{i, w_i \neq 0\}|$.

The sparesest solution $\hat{v}$ of (12) can be found by solving Basis Pursuit (BP) problem $P_1$ [16]:

$$\hat{w} = \arg\min_{w \in \mathbb{R}^n} \|w\|_{l_0} \text{ subject to } u = Aw, \qquad (14)$$

in which, $l_1$ norm is $\|w\|_{l_1} := \sum_{i=1}^{n}|w_i|$.

The above-described problem assumes that we are given the exact form of the reconstructed signal. This is rarely the case in practice, because the measurements are often affected by noise. To reconstruct the signal in case of noise-affected measurements, we have:

$$u = Aw + e, \quad (15)$$

in which, $e$ represents the noise $\|e\|_{l_2} \leq \varepsilon$, $P_1$ problem can be rewritten as follows [16]:

$$\hat{w} = \arg\min_{w \in R^n} \|w\|_{l_1} \text{ subject to } u = \|u - Av\|_{l_2} \leq \varepsilon \quad (16)$$

In particular form, $P_1$ problem for DBIM method is written by

$$\Delta \bar{T} = \arg\min_{\Delta \bar{T}} \|\Delta \bar{p}^{sc}{}_t - \overline{M_t} \Delta \bar{T}\|_2^2 + \zeta \|\Delta \bar{T}\|_1 \quad (17)$$

where $\zeta$ is the regularization parameter in $P_1$ problem.

**3.2 Compressive Sampling using Chaos Filters**

In this sub-section, we consider a very simple chaotic sequence, the Logistic map, and its transformed version to have Gaussian-like behavior.

In conventional compressive sampling technique, the measurement matrix $\Phi$ is random. It is well-known that the hardware implementation of a deterministic system is commonly simpler than that of a random one. Therefore, matrix $\Phi$ is made to be chaotic. This can be generated from a deterministic system. Thanks to the design of the random filter for CS in [17], a design of a chaotic filter for CS was proposed in [18], where a chaotic $\Phi$ is constructed from a chaotic sequence $h_G(n)$. This sequence is obtained by first generating the Logistic map

$$q(n + 1) = \rho q(n)(1 - q(n)), \quad (18)$$

and then converting it by the Logit Transform to be Gaussian-like as

$$q_G(n) = \ln\left[\frac{q(n)}{1 - q(n)}\right] \quad (19)$$

Because of concerning $q(n)$ to be chaotic, the control parameter $\rho$ must be equal to 4. The initial condition $q(0)$ is very sensitive in the sense that the output chaotic sequence is completely different for a small change of $q(0)$. More detail construction $\Phi$ from $q_G(n)$ can be found in [18]. And then, the image reconstruction is performed using the Orthogonal Matching Pursuit technique.

## IV. THE PPROPOSED METHOD

The complexity of the reconstruction system depends on the total number of iterations ($N_{sum}$), the number of transmitters ($N_t$) and receivers ($N_r$).

DBIM uses Born approximation to compute iterative solutions of a nonlinear inverse scattering problem. The Tikhonov regularization problem can be resolved directly or indirectly using an iterative method. However, the iterative method is more efficient than the direct one, especially when M is sparse or has a special form (e. g., wavelet matrices or partial Fourier). In [10], M is determined by using multiple transmitters and detectors placed at equal distances as illustrated in Fig. 5. This configuration would make M become large, thus, it is not efficient for the iteration steps.

In this paper, we propose to use a chaotic under-sampling configuration of detectors, as shown in Fig. 6, with the number of detectors is smaller than that in the conventional configuration. With a reduced number of measurements (i. e., the size of M), and hence reduced the computational complexity in the iteration process, the proposed configuration maintains a quality of the reconstruction comparable to that obtained by the conventional configuration. Note that the transmitters are still placed at equal distance as in the conventional configuration.

The undersampling ratio is defined as follows:

$$r = \frac{N_T N_R}{N^2} \quad (20)$$

When $N_T = N_R = N^2$, $r = 1$; this corresponds to the conventional configuration with full linear sampling. Otherwise, we have $r < 1$ and this corresponds to the undersampling configuration. In practice, the value of maximum number of measurements depends on the accuracy of the mechanical system rotating around the object, which assembles the transmitters and detectors.

The implementation process of the conventional method is shown in Fig. 2. The input is here the ideal target function and the output is the reconstructed target function. In this method, the measurement configuration of linear transmitter-and-detector locations is used, and then the ideal target function of interest is reconstructed using Tikhonov regularization.

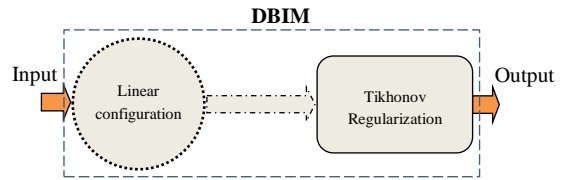

Fig. 2. The implementation process of the conventional method

The implementation process of the proposed method is shown in Fig. 3. In this method, the measurement configuration of transmitters and detectors with linear transmitter locations and chaotic detector locations is used, and then the ideal target function of interest is reconstructed using $L_1$ regularization [19].

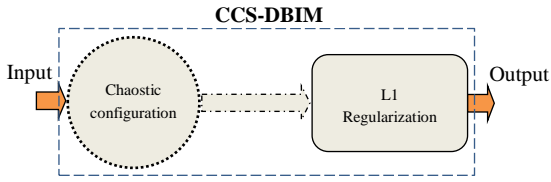

Fig. 3. The implementation process of the proposed method

The CCS-DBIM procedure is presented in Algorithm 2.

**Algorithm 2.** The Chaos Compressive Sampling DBIM – CCS-DBIM

Set up the measurement configuration of linear tranmitters and chaotic receivers locations
Choose initial values: $\bar{T}_{(n)} = \bar{T}_{(0)}$ and $\bar{p}_0 = \bar{p}^{inc}$ using (21)
**For** $n = 1\ to\ N_{DBIM}$, **do**
1. Calculate $\bar{B}$ and $\bar{C}$
2. Calculate $p, \bar{p}^{sc}$ corresponding to $\bar{T}_{(n)}$ using (5, 6)
3. Calculate $\Delta \bar{p}^{sc}$ using (7)
4. Calculate $\Delta \bar{T}_{(n)}$ using $L_1$ regulation (17)
5. Calculate $\bar{T}_{(n+1)} = \bar{T}_{(n)} + \Delta \bar{T}_{(n)}$
**End For**

## V. SIMULATION AND RESULTS

Simulation parameters: Frequency f = 1 MHz; $N_{sum}$ = 8; N = 21 (i.e. Number of variables is $N^2$ = 21×21 = 441); Scattering area diameter = 7.3 mm; Sound contrast 5%; Gaussian noise 10%; Distances from transmitters and receivers to the center of the target are 100 mm.

The incident pressure for a Bessel beam of zero order in two-dimensional case is

$$\bar{p}^{inc} = J_0(k_0|r - r_k|), \quad (21)$$

where $J_0$ is the 0th order Bessel function and $|r - r_k|$ is the distance between the transmitter and the k$^{th}$ point in the ROI.

Fig. 4 shows the ideal target function T(r) (4). The target is placed at the center of the meshing area.

Fig. 5a shows the conventional configuration of transmitters and detectors using linear transmitter-and-detector locations in case of $N_T = N_R$ = 22. Fig. 5b shows the histogram of linear detector locations over full circle in case of $N_R$ = 22.

Fig. 6a shows the proposed configuration of transmitters and detectors with linear transmitter locations and chaotic detector locations in case of $N_T = N_R$ = 16. Fig. 6b shows the histogram of chaotic detector locations over full circle in case of $N_R$ = 16.

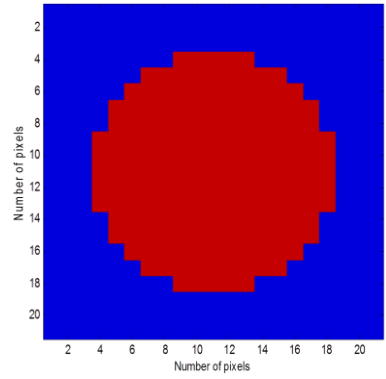

Fig. 4. Ideal target function (N = 21)

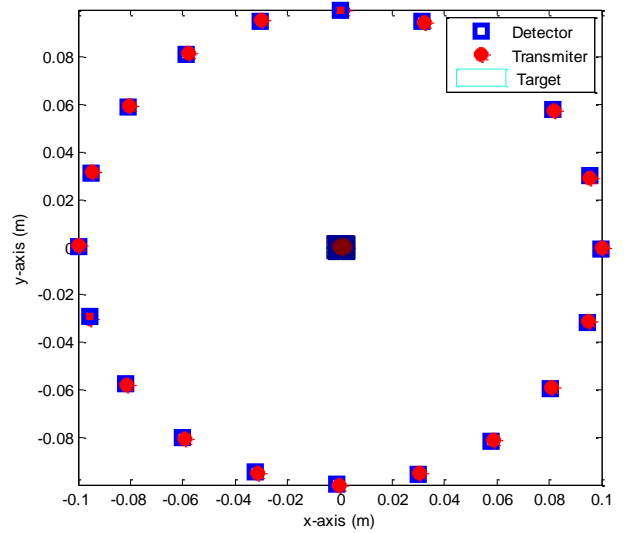

Fig. 5a. Conventional configuration of transmitters and detectors using linear transmitter-and-detector locations ($N_T = N_R$ = 20, r = 0.826)

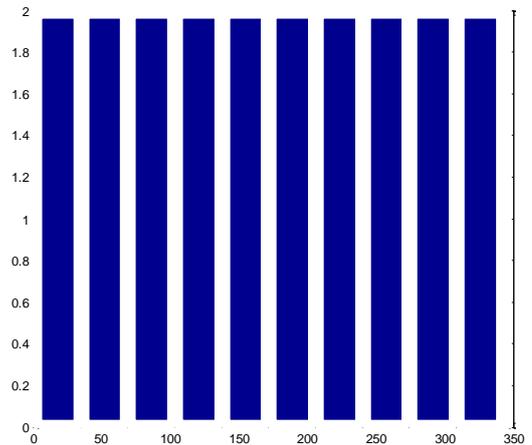

Fig. 5b. Histogram of linear detector locations over full circle ($N_R$ = 20)

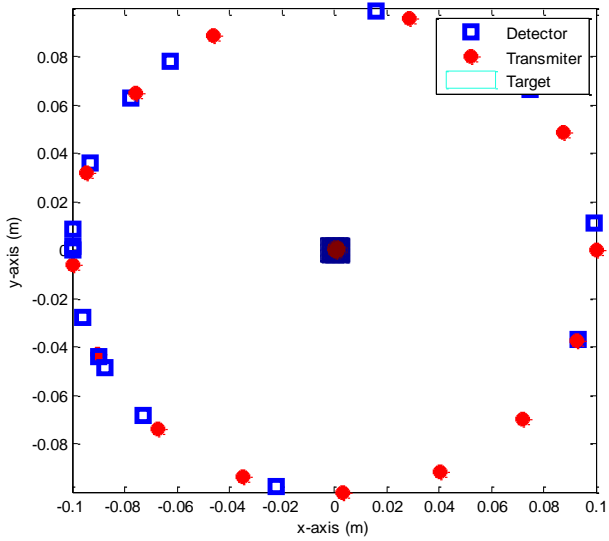

Fig. 6a. Proposed configuration of transmitters and detectors using linear transmitter locations and chaotic detector locations ($N_T = N_R = 16$, r = 0.581)

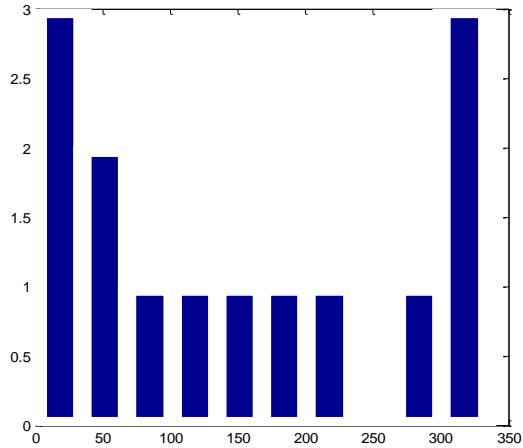

Fig. 6b. Histogram of chaotic detector locations over full circle ($N_R = 16$)

To quantify the efficiency of the proposed approach, we acquire the target functions for a series of iterations. Then, the error in the reconstructed image is determined and compared to the original image in each iteration. Suppose that $m$ is a $P \times Q$ original image (i.e. ideal target function) and $\hat{m}$ is the reconstructed image. The error can be defined as:

$$\varepsilon = \frac{1}{P \times Q} \sum_{i=1}^{P} \sum_{j=1}^{Q} \frac{|m_{ij} - \hat{m}_{ij}|}{|m_{ij}|} \tag{22}$$

Table 1 shows the normalized errors and runtimes of the DBIM and CCS-DBIM methods through iterations with different number of transmitters ($N_T$) and receivers ($N_T$).

Firstly, in term of the runtime after $N_{sum}$ iterations, the simulation results indicate that the runtime of the CCS-DBIM method is significantly larger than that of the DBIM method.

It is clearly shown in Table 2, the minimum and maximum increased percent of runtime, compared to the conventional method, when using the proposed method, are 19.13% (in case of 900 measurements) and 69.17% (in case of 324 measurements), respectively.

Secondly, in term of the normalized error after $N_{sum}$ iterations, the simulation results indicate that the image reconstruction quality of the CCS-DBIM method is worse than the DBIM method when $r < 0.5$ and is significantly better than the one of the conventional method when $r > 0.5$. In case of $r < 0.5$, although the reconstruction quality of the proposed method is not as good as the conventional method, it can successfully reconstruct the target function when $r$ is very small (in case of r = 0.082 and 0.145). Meanwhile, the conventional method cannot reconstruct the target function (i.e. NaN in Table 1). In case of $r > 0.5$, the reconstruction quality of the proposed method is significantly better than the conventional method. It is clearly shown in Table 3, the minimum and maximum decreased percent of normalized error, compared to the conventional method, when using the proposed method, are 26.21% (in case of 484 measurements) and 96.52% (in case of 900 measurements), respectively. However, in practice, we concern the case that it offers the best performance with a small number of measurements. Therefore, we are interested in the case of $r = 0.735$ (i.e. 324 measurements), it offers the 90.72% reduced normalized error (as shown in Fig. 7), compared to the conventional method. In general, the simulation results have demonstrated that the CCS-DBIM method is a very robust tool for a very high-quality reconstruction. It would be a very promising approach in practical applications of modern biomedical imaging technology.

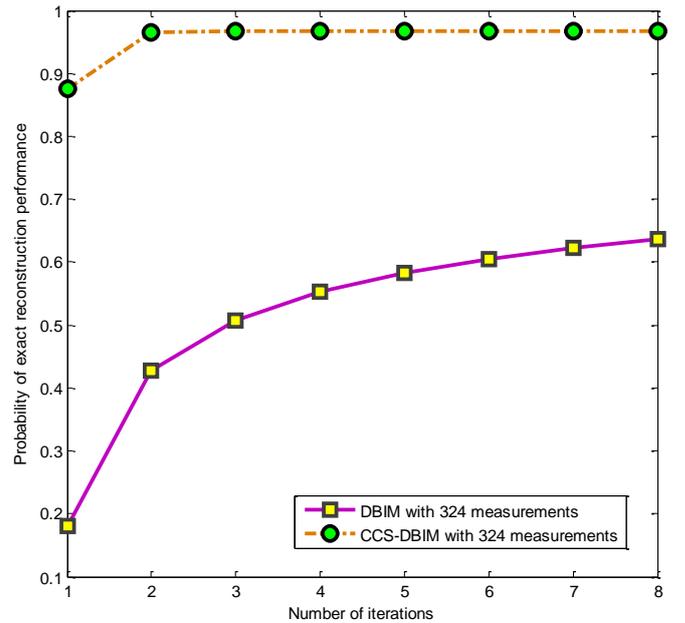

Fig. 7. Probability of exact reconstruction performance comparison of the conventional and proposed methods

Fig. 8a presents the error performance of the CCS-DBIM method (in case of $N_T = N_R = 16$, i.e. number of measurements = 16x16 = 256) in comparison with the conventional DBIM one (in case of $N_T = N_R = 22$, i.e. number of measurements = 22x22 = 484). Although the number of measurements of the CCS-DBIM method is approximately half the one of the DBIM method, both methods offer the same image reconstruction quality after the sixth iteration step. With the same normalized error, in the CCS-DBIM method, we only need 3 iterations, meanwhile, in the DBIM method, we need 6 iterations. Therefore, in this scenario, when using the proposed method, we save half of number of measurements and iterations. It is also shown the high performance of the proposed method (with 400 measurements), compared to the conventional method (with 900 measurements) in Fig. 8b. However, the only disadvantage of the proposed method is that the runtime of this method is dramatically longer than the conventional method.

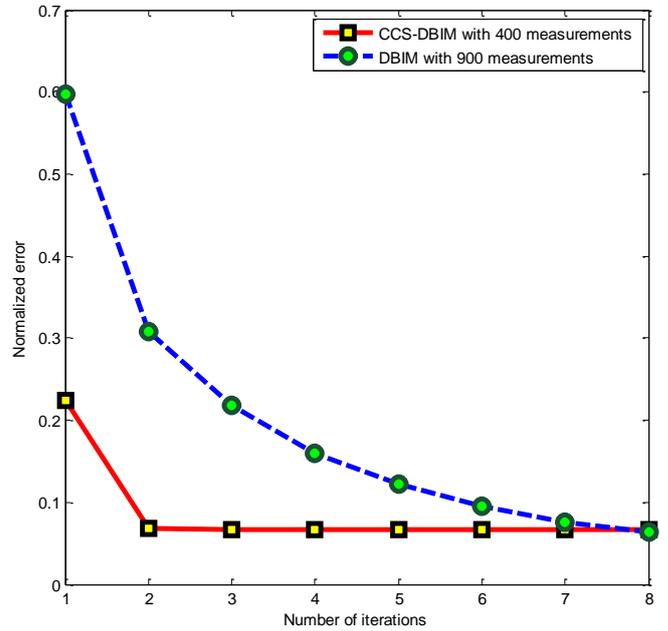

Fig. 8b. Normalized error comparison of the (900 measurements) conventional and (400 measurements) proposed methods

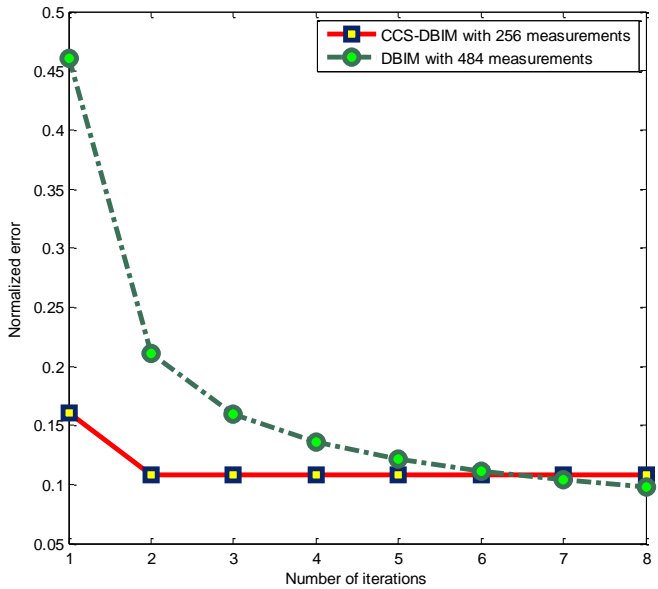

Fig. 8a. Normalized error comparison of the (484 measurements) conventional and (256 measurements) proposed methods

## VI. CONCLUSIONS

Based on inverse scattering theory, the DBIM is a well-known quantitative imaging approach for detecting very small targets thanks to its mechanical properties. Chaotic compressive sampling technique is a promising approach for feasible hardware implementation in practical applications. This paper has successfully applied CCS technique for setting up the measurement configuration for the DBIM, and then the target is reconstructed using $L_1$ least square problem in order to improve the quality of the image reconstruction. This method also offers a very simple setting compared to the others. Simulation scenarios of sound contrast reconstruction were implemented to demonstrate the very good performance of this method. The scheme can be further developed by 3D reconstruction and experiment.

ACKNOWLEDGMENT

The work was supported by VNU Development Fund of Science and Technology, VNU Scientists Club (VSL-VNU Scientific Links).

Table 1. The normalized errors and runtimes of the DBIM and CCS-DBIM methods through iterations with different $N_T$ and $N_R$

| Number of Transmitters ($N_T$) and Receivers ($N_R$) | Methods | Normalized error from the first iteration to the eighth iteration | | | | | | | | Runtime (seconds) |
|---|---|---|---|---|---|---|---|---|---|---|
| $N_T = N_R = 6$ | DBIM | 0.8682 | 0.8438 | NaN | NaN | NaN | NaN | NaN | NaN | 69.692787 |
| (r = 0.082) | CCS-DBIM | 1.2077 | 1.2102 | 1.2105 | 1.2105 | 1.2105 | 1.2105 | 1.2105 | 1.2105 | 42.139530 |
| $N_T = N_R = 8$ | DBIM | 0.7964 | 0.7515 | 0.7490 | 0.7489 | 0.7489 | NaN | NaN | NaN | 61.308944 |
| (r = 0.145) | CCS-DBIM | 1.1587 | 1.1718 | 1.1721 | 1.1721 | 1.1721 | 1.1721 | 1.1721 | 1.1721 | 47.157182 |
| $N_T = N_R = 10$ | DBIM | 0.7305 | 0.6811 | 0.6779 | 0.6772 | 0.6771 | 0.6770 | 0.6770 | 0.6770 | 40.428921 |
| (r = 0.227) | CCS-DBIM | 1.1123 | 1.1224 | 1.1226 | 1.1226 | 1.1226 | 1.1226 | 1.1226 | 1.1226 | 53.017146 |
| $N_T = N_R = 12$ | DBIM | 0.6808 | 0.6140 | 0.6083 | 0.6073 | 0.6070 | 0.6069 | 0.6069 | 0.6069 | 48.419012 |
| (r = 0.327) | CCS-DBIM | 0.8834 | 0.8945 | 0.8950 | 0.8950 | 0.8950 | 0.8950 | 0.8950 | 0.8950 | 63.057426 |
| $N_T = N_R = 14$ | DBIM | 0.9367 | 0.6457 | 0.5824 | 0.5547 | 0.5398 | 0.5308 | 0.5254 | 0.5218 | 55.754608 |
| (r = 0.444) | CCS-DBIM | 0.7025 | 0.7084 | 0.7085 | 0.7085 | 0.7085 | 0.7085 | 0.7085 | 0.7085 | 90.422976 |
| **$N_T = N_R = 16$** | DBIM | 0.5272 | 0.4629 | 0.4585 | 0.4576 | 0.4572 | 0.4570 | 0.4570 | 0.4570 | 66.137258 |
| **(r = 0.581)** | **CCS-DBIM** | **0.1604** | **0.1078** | **0.1082** | **0.1082** | **0.1082** | **0.1082** | **0.1082** | **0.1082** | **185.779095** |
| $N_T = N_R = 18$ | DBIM | 0.8196 | 0.5734 | 0.4919 | 0.4480 | 0.4176 | 0.3948 | 0.3773 | 0.3632 | 77.524221 |
| (r = 0.735) | CCS-DBIM | 0.1240 | 0.0342 | 0.0338 | 0.0337 | 0.0337 | 0.0337 | 0.0337 | 0.0337 | 251.473327 |
| $N_T = N_R = 20$ | DBIM | 0.4749 | 0.2760 | 0.2356 | 0.2225 | 0.2158 | 0.2115 | 0.2086 | 0.2066 | 94.343578 |
| (r = 0.907) | CCS-DBIM | 0.2243 | 0.0689 | 0.0672 | 0.0670 | 0.0670 | 0.0669 | 0.0668 | 0.0668 | 237.118941 |
| $N_T = N_R = 22$ | **DBIM** | **0.4604** | **0.2106** | **0.1598** | **0.1353** | **0.1209** | **0.1110** | **0.1036** | **0.0973** | **112.128716** |
| (r = 1.098) | CCS-DBIM | 0.3255 | 0.0777 | 0.0729 | 0.0724 | 0.0723 | 0.0721 | 0.0719 | 0.0718 | 234.584982 |
| $N_T = N_R = 24$ | DBIM | 0.5754 | 0.2832 | 0.1321 | 0.0942 | 0.0725 | 0.0641 | 0.0534 | 0.0632 | 125.724742 |
| (r = 1.306) | CCS-DBIM | 0.3870 | 0.0310 | 0.0197 | 0.0192 | 0.0190 | 0.0188 | 0.0186 | 0.0184 | 225.159681 |
| $N_T = N_R = 26$ | DBIM | 0.5545 | 0.1933 | 0.1141 | 0.0846 | 0.0685 | 0.0585 | 0.0516 | 0.0464 | 144.661175 |
| (r = 1.533) | CCS-DBIM | 0.1768 | 0.0129 | 0.0066 | 0.0064 | 0.0064 | 0.0064 | 0.0064 | 0.0064 | 201.250795 |
| $N_T = N_R = 28$ | DBIM | 0.4905 | 0.1680 | 0.0858 | 0.0570 | 0.0417 | 0.0329 | 0.0271 | 0.0229 | 170.032561 |
| (r = 1.778) | CCS-DBIM | 0.1338 | 0.0080 | 0.0031 | 0.0030 | 0.0029 | 0.0029 | 0.0029 | 0.0029 | 238.488285 |
| $N_T = N_R = 30$ | DBIM | 0.5971 | 0.3079 | 0.2179 | 0.1602 | 0.1215 | 0.0948 | 0.0764 | 0.0633 | 212.923908 |
| (r = 2.041) | CCS-DBIM | 0.1466 | 0.0062 | 0.0022 | 0.0022 | 0.0022 | 0.0022 | 0.0022 | 0.0022 | 263.288455 |

Table 2. The runtime of the DBIM and CCS-DBIM methods after the eighth iteration with different measurements

| Number of measurements | 100 | 144 | 196 | 256 | 324 | 400 | 484 | 576 | 676 | 784 | 900 |
|---|---|---|---|---|---|---|---|---|---|---|---|
| Runtime of DBIM (s) | 40.428921 | 48.419012 | 55.754608 | 66.137258 | 77.524221 | 94.343578 | 112.128716 | 125.724742 | 144.661175 | 170.032561 | 212.923908 |
| Runtime of CCS-DBIM (s) | 53.017146 | 63.057426 | 90.422976 | 185.779095 | 251.473327 | 237.118941 | 234.584982 | 225.159681 | 201.250795 | 238.488285 | 263.288455 |
| Increased percent of runtime | 23.74% | 23.21% | 38.34% | 64.40% | **69.17%** | 60.21% | 52.20% | 44.16% | 28.12% | 28.70% | **19.13%** |

Table 3. The normalized error of the DBIM and CCS-DBIM methods after the eighth iteration with different measurements

| Number of measurements | 100 | 144 | 196 | 256 | 324 | 400 | 484 | 576 | 676 | 784 | 900 |
|---|---|---|---|---|---|---|---|---|---|---|---|
| Error of DBIM | 0.6770 | 0.6069 | 0.5218 | 0.4570 | 0.3632 | 0.2066 | 0.0973 | 0.0632 | 0.0464 | 0.0229 | 0.0633 |
| Error of CCS-DBIM | 1.1226 | 0.8950 | 0.7085 | 0.1082 | 0.0337 | 0.0668 | 0.0718 | 0.0184 | 0.0064 | 0.0029 | 0.0022 |
| Decreased percent of error | 39.69% | 32.19% | 26.35% | 76.32% | 90.72% | 67.67% | **26.21%** | 70.89% | 86.21% | 87.34% | **96.52%** |

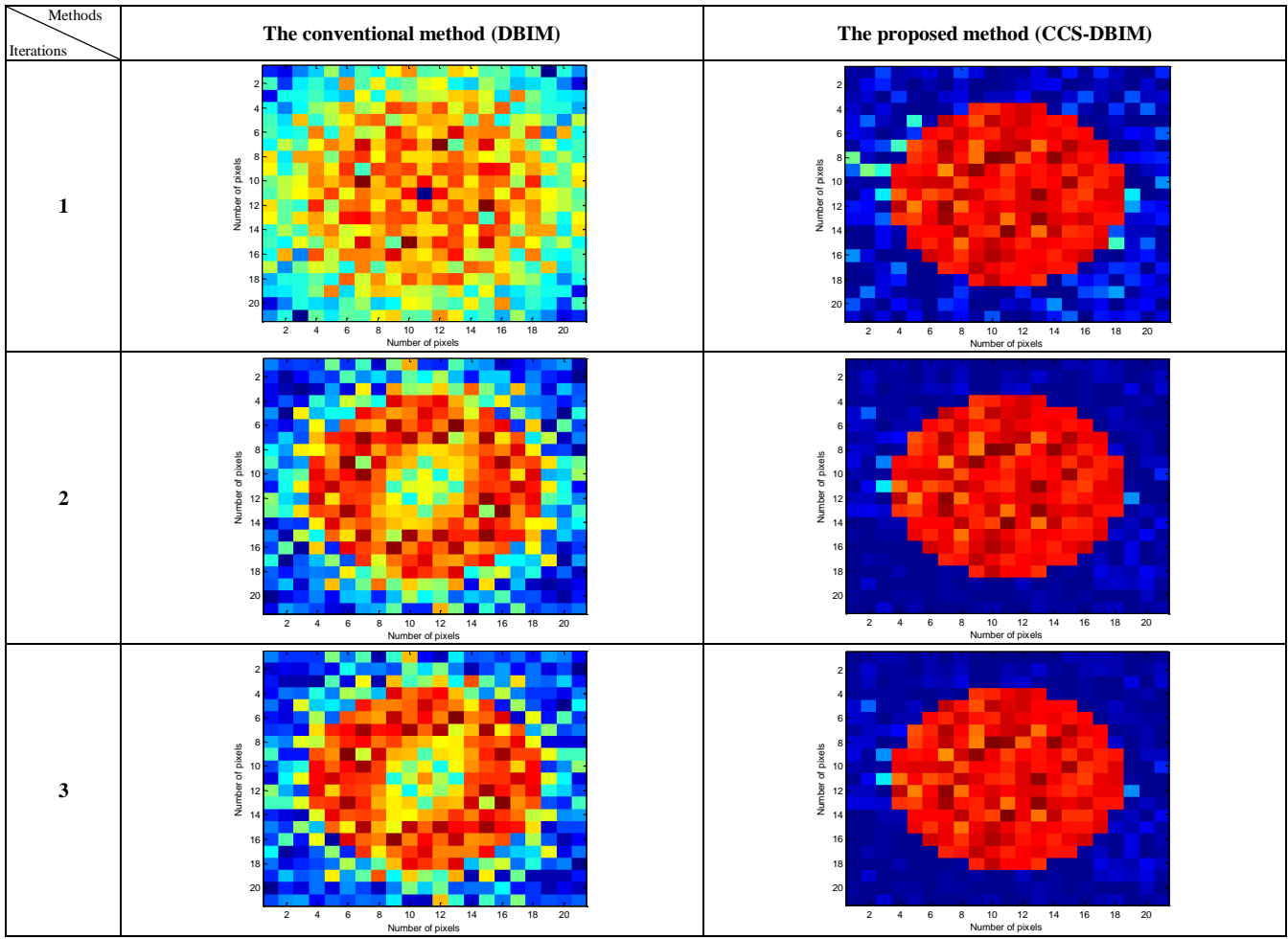

| Methods / Iterations | The conventional method (DBIM) | The proposed method (CCS-DBIM) |
|---|---|---|
| **1** | | |
| **2** | | |
| **3** | | |

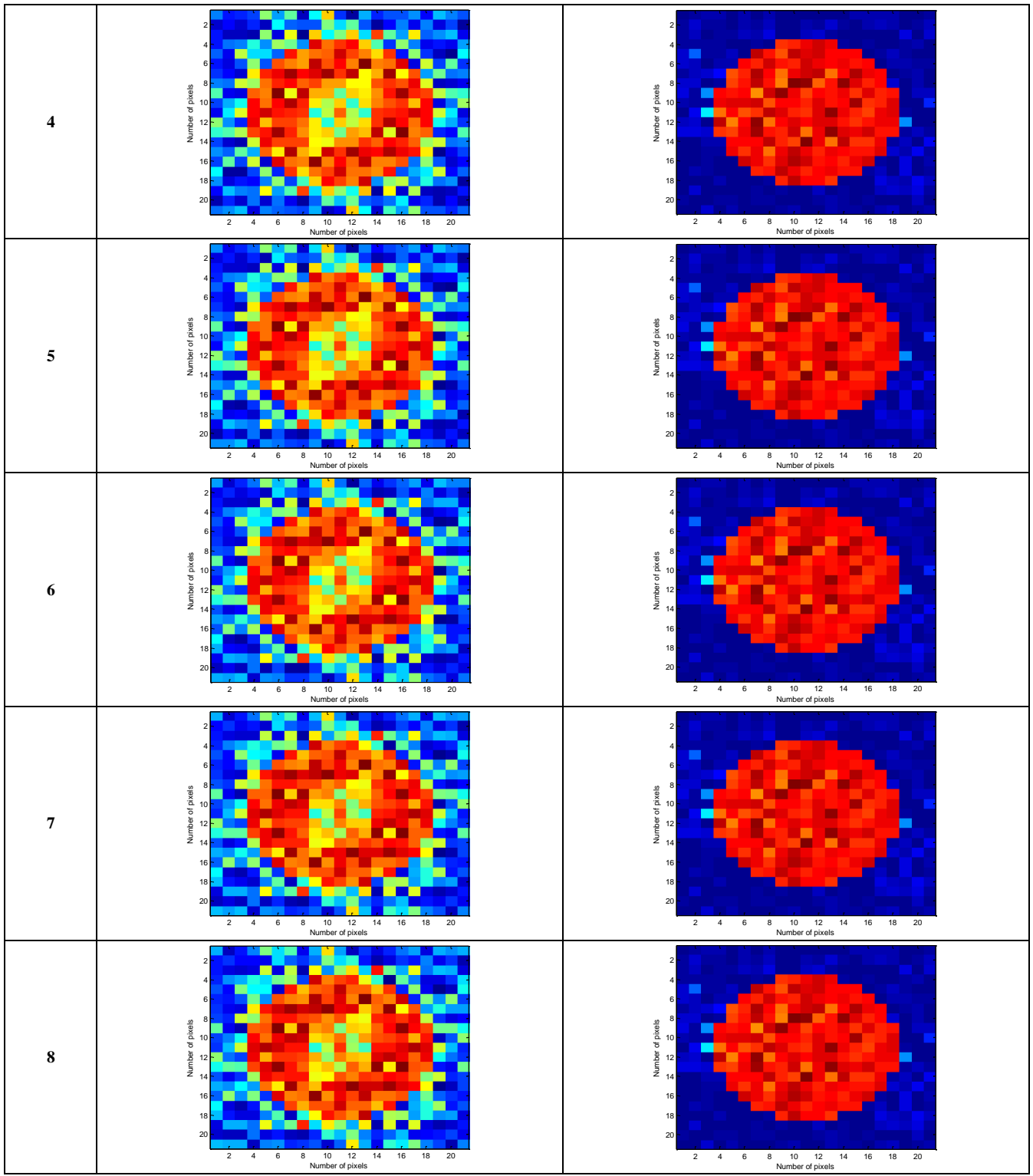

Fig. 9. The reconstructed results of the DBIM and CCS-DBIM methods through iterations in case of $N_T = N_R = 16$, $r = 0.581$